\documentclass{IEEEcsmag}

\usepackage[colorlinks,urlcolor=blue,linkcolor=blue,citecolor=blue]{hyperref}
\expandafter\def\expandafter\UrlBreaks\expandafter{\UrlBreaks\do\/\do\*\do\-\do\~\do\'\do\"\do\-}
\usepackage{upmath}

\usepackage{multirow}
\usepackage{cite}
\usepackage{array}
\usepackage[table,dvipsnames]{xcolor}
\definecolor{IEEEblue}{rgb}{0.0, 0.53, 0.74}
\newcolumntype{M}[1]{>{\centering\arraybackslash}m{#1}}
\newcolumntype{B}[1]{>{\centering\arraybackslash\columncolor{IEEEblue!40}}m{#1}}
\usepackage{tabularx}
\usepackage{xspace}
\usepackage{graphicx}

\graphicspath{{figs/}}

\newcommand{\etal}       {\textit{et~al.}\xspace}

\begin{document}

\title{EDHOC is a New Security Handshake Standard: An Overview of Security Analysis}

\author{Elsa López Pérez}
\affil{Inria}

\author{Göran Selander}
\affil{Ericsson}

\author{John Preuß Mattsson}
\affil{Ericsson}

\author{Thomas Watteyne}
\affil{Inria}

\author{Mališa Vučinić}
\affil{Inria}


\begin{abstract}
The paper wraps up the call for formal analysis of the new security handshake protocol EDHOC by providing
    an overview of the protocol as it was standardized,
    a summary of the formal security analyses conducted by the community, and
    a discussion on open venues for future work.
\end{abstract}

\maketitle

\begin{IEEEkeywords}
    Standardization,
    Lightweight Security,
    Wireless Communication,
    Protocol Design and Analysis.
\end{IEEEkeywords}

\section{Introduction}

The rise of the Internet of Things (IoT) has driven organizations like the Internet Engineering Task Force (IETF) to develop protocols that meet the requirements of the involved devices and networks. 
Some of the challenges are their low processing power, scarce bandwidth, battery lifetimes and reduced data rates.
To address these issues, the Internet community has developed and standardized protocols that are tailored for constrained environments.
The results of these efforts include
    the Constrained Application Protocol (CoAP) and
    the Object Security for Constrained RESTful Environments (OSCORE).
CoAP is a specialized web transfer protocol that provides the REST services of HTTP but with reduced overhead and processing.
OSCORE is a security protocol that can be applied to protect CoAP communication, 
including end-to-end encryption and integrity across CoAP proxies, replay protection 
and binding of responses to requests.


OSCORE itself does not define a key establishment protocol. 
Prior to using OSCORE, the communicating parties must establish a security association, including a shared cryptographic key through some out-of-band mechanism.
To resolve this matter, the IETF created the Lightweight Authenticated Key Exchange (LAKE) working group which developed and standarized the Ephemeral Diffie-Hellman Over COSE (EDHOC) protocol.
EDHOC is designed to enable an authenticated Diffie-Hellman key exchange and shared secret key derivation between two peers, both possibly operating on resource-constrained devices utilizing low-power IoT radio communication technologies.
EDHOC, like OSCORE, builds on the Concise Binary Object Representation (CBOR) 
encoding and the object security format CBOR Object Signing and Encryption (COSE), enabling reduced message sizes and combined code footprint.


EDHOC and OSCORE provide an application layer alternative or complement to Transport Layer Security (TLS) for protection of CoAP.
TLS 1.3 is the go-to standard for web security but faces challenges in constrained environments due to large handshake message sizes and more elaborate state machines.
The same applies to Datagram TLS (DTLS) 1.3 used for connectionless transports, such as User Datagram Protocol (UDP).
A DTLS handshake transfers around 1~kB of data \cite{draft-ietf-iotops-security-protocol-comparison-06}.

EDHOC, for comparison, allows for handshakes that transfer 100$+$~bytes of data, requires only three mandatory flights, is transport agnostic, and code size can be kept low by reusing the same elements as OSCORE.
Fedrecheski~\etal~\cite{fedrecheski24performance} show that EDHOC achieves a 7.75$\times$ reduction in message footprint, 1.9$\times$ reduction in energy and time, and uses up to 4$\times$ less flash and RAM than DTLS~1.3.


In November 2021, Vučinić~\etal~\cite{vucinic22lightweight} invited the formal analysis community to study the EDHOC protocol. 
During the following six-month period, the standardization process was ``frozen'' and no modifications to the protocol were done. 
This paper wraps up the formal analysis stage by offering an overview of the protocol as formalized in RFC 9528 and RFC 9529 and a summary of the security analyses conducted by the time of publication. 
The paper also identifies potential areas for improvement and outlines future research directions.

\section{Security Goals}
\label{sec:Security_goals}


The security goals of the protocol adhere to the requirements established by Vučinić~\etal~\cite{draft-ietf-lake-reqs-04}.
They can be summarized as follows:
\begin{itemize}
    \item \textbf{Confidentiality.}
        The shared secret established at the end of the session must be known only to the two authenticated peers.
        In addition, an active attacker who has compromised either one of the peer's private keys shall still not be able to compute past session keys (Forward Secrecy).
        This is achieved by generating session keys from ephemeral keys, which are freshly defined in each session.
        Therefore, even if a session key is compromised, only the data from the current session is at risk, but not past communications.
    \item \textbf{Mutual authentication.}
        At the end of the session, each peer should have freshly authenticated the other peer.
        Both peers must agree on a fresh session identifier, roles and credentials. 
        Compromising the long-term secret of one party should not break that party's authentication of their peer in the given session (Key Compromise Impersonation resistance). 
        In addition, the protocol shall offer protection against identity misbinding attacks, where a peer becomes unknowingly associated with a third party.    
    \item \textbf{Identity protection.}
        ``Identity'' refers to a unique identifier that allows different entities within the protocol to recognize and authenticate each other.
        Identity may be represented by a cryptographic certificate, public keys, MAC addresses or any other unique identifier exchanged during the protocol execution. 
        The protocol must protect the identity of one of the peers against active attackers and the identity of the other peer against passive attackers.
    \item \textbf{Cryptographic strength.}
        The target security level of the protocol's key shall be greater or equal than 128~bits, meaning that the complexity of an attack is greater or equal to $2^{128}$ to brute-force the key.
        This security level targets the strength of the authentication, established keys, and protection of negotiation for all cryptographic parameters.
    \item \textbf{Protection of external security data.}
        For efficiency, the protocol should support integration of external security applications using so called ``External Authorization Data'' (EAD) message fields having the same level of protection as the protocol message they are carried within.
    \item \textbf{Downgrade protection.}
        In response to long deployment lifetimes, the protocol must support cryptographic agility, including modularity and negotiation of preferred cryptographic primitives.
        At the end of the session, both peers must agree on the cryptographic algorithms that were proposed and chosen.
        Downgrade protection is crucial to prevent attackers from forcing the use of weaker security features.
        Some related attacks include
            \textit{(1)} cipher suite downgrade attack, in which attackers attempt to manipulate the negotiation process to force the use of less secure cipher suites, or
            \textit{(2)} key material downgrade attack, in which an adversary intercepts the key derivation process and modifies the generated key to use weaker or compromised key material.
\end{itemize}

\section{The EDHOC Protocol}

\paragraph{A Primer on SIGMA}


\begin{figure*}
    \centering
    \includegraphics[width=0.75\textwidth]{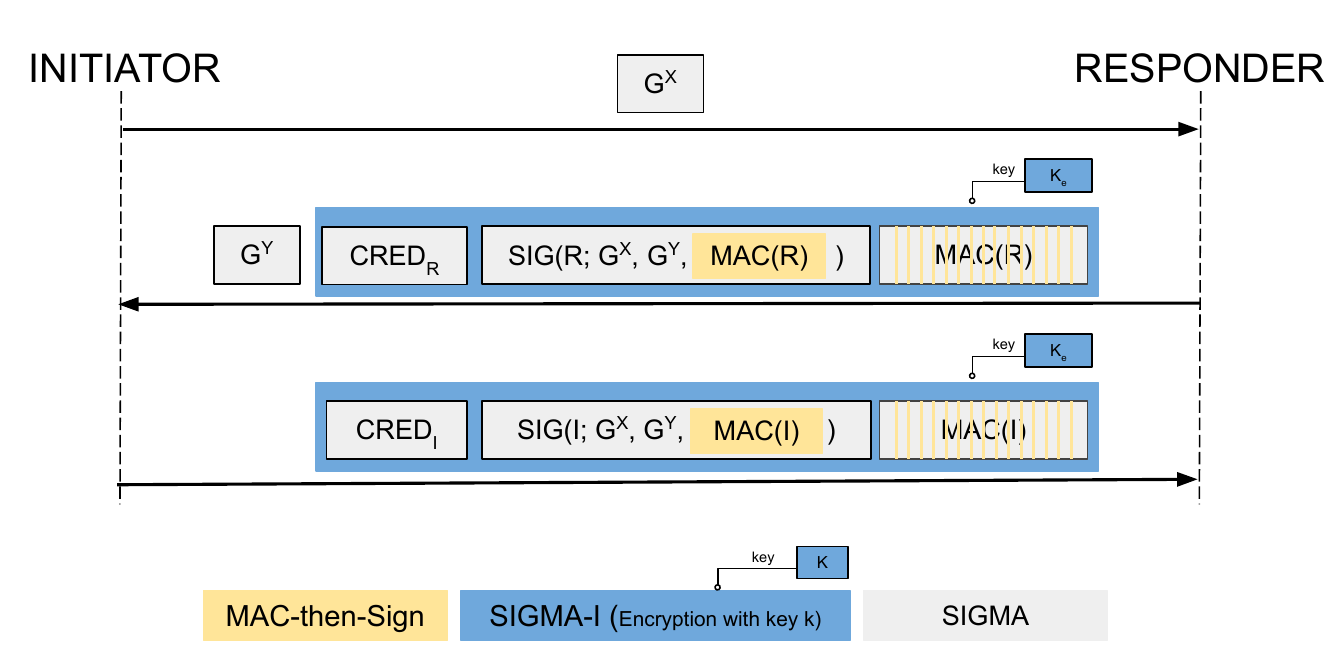}
    \caption{%
        Message flow of the SIGMA protocol.
        The blue boxes denote additions to the SIGMA-I variant.
        The yellow boxes denote additions to the MAC-then-Sign variant.
    }
    \label{fig:SIGMA}
\end{figure*}

EDHOC is designed following the SIGMA protocol, a family of key-exchange protocols that introduce a general approach to building authenticated Diffie-Hellman (DH) protocols using a combination of digital signatures and message authentication code (MAC) functions. 
SIGMA is widely used and constitutes the cryptographic basis of known protocols such as the Internet Key Exchange (IKE) (IKE version 1 and version 2) or Transport Layer Security (TLS) version~1.3.
In SIGMA, each party can authenticate to the other without needing to know the peer's identity beforehand.
This property allows the protocol to support identity protection, a requirement that plays a central role in its design.
More specifically, it decouples the authentication of the ephemeral Diffie-Hellman public key ($G^{X}$, $G^{Y}$) performed via digital signatures, from the binding of key and identities, done by computing a MAC function.
The most basic form of the SIGMA protocol, without identity protection, consists in sending both the digital signatures and the MAC (see Fig.~\ref{fig:SIGMA}).
The output of the protocol is a session key derived from the Diffie-Hellman shared secret ($G^{XY}$).


In case the identity protection functionality is needed, one of the peers can delay communicating its own identity until it learns the peer’s identity in an authenticated form. 
This gives rise to two variants of SIGMA:
    \textit{(1)} SIGMA-I, which protects the identity of the peer initiating the session, called Initiator, and
    \textit{(2)} SIGMA-R, which protects the identity of the peer engaging in an already initiated session, called Responder. 


To reduce the size of the messages on the wire, of special interest in constrained environments, SIGMA introduces the MAC-then-Sign variant.
In MAC-then-Sign, MAC is included under the Signature, allowing to reduce message length by not sending the MAC explicitly.
The MAC-then-Sign messages are therefore shorter for the length of the MAC.
As long as the MAC covers the identity of the signer the same security of the basic SIGMA protocol is preserved. 
Fig.~\ref{fig:SIGMA} shows the differences between the basic SIGMA protocol, SIGMA-I and MAC-then-Sign variants.

\paragraph{EDHOC Additions to SIGMA}


\begin{table}
    \centering
    \rowcolors{2}{IEEEblue!10}{white}
    \begin{tabular}{c|c|c}
        \rowcolor{IEEEblue!40}
        ID & Initiator             & Responder             \\[5pt]
        \hline
        0  & Signature             & Signature             \\[5pt]
        1  & Signature             & Static Diffie-Hellman \\[5pt]
        2  & Static Diffie-Hellman & Signature             \\[5pt]
        3  & Static Diffie-Hellman & Static Diffie-Hellman \\[5pt]
    \end{tabular}
    \caption{Current authentication methods registered by IANA.}
    \label{tab:method_EDHOC}
\end{table}

EDHOC cryptographic core follows the MAC-then-Sign variant of the SIGMA-I protocol.
However, apart from conventional signature keys used for authentication, EDHOC enables the use of static Diffie-Hellman keys.
The EDHOC ``authentication method'', sent in the first message, defines which type of the authentication key the peers are using (see Table~\ref{tab:method_EDHOC}).


The use of static Diffie-Hellman keys allows significant reduction in message size.
While following the SIGMA-I MAC-then-Sign structure, EDHOC replaces the digital signature with a MAC in case the peer is using the Static Diffie-Hellman key for its authentication.
This works because the peer can combine the static Diffie-Hellman key pair with an ephemeral key pair of the other peer and produce an ephemeral-static shared secret to compute the MAC.
The peer then needs to send a shorter MAC (e.g. 8~bytes) instead of a digital signature (e.g.~64~bytes).

\begin{table*}
    \centering
    \rowcolors{2}{IEEEblue!10}{white} 
    \begin{tabular}{M{1.5cm}|M{3cm}|M{1.5cm}|M{1.5cm}|M{1.7cm}|M{1.5cm}|M{3cm}}
        \rowcolor{IEEEblue!40}
        ID                 & AEAD               & Hash     & MAC length & ECDH Curve & Signature & Application AEAD \\[5pt]
        \hline
        0                  & AES-CCM-16-64-128  & SHA-256  & 8          & X25519     & EdDSA     & AES-CCM-16-64-128 \\[5pt]
        1                  & AES-CCM-16-128-128 & SHA-256  & 16         & X25519     & EdDSA     & AES-CCM-16-64-128 \\[5pt]
        2                  & AES-CCM-16-64-128  & SHA-256  & 8          & P-256      & ES256     & AES-CCM-16-64-128 \\[5pt] 
        3                  & AES-CCM-16-128-128 & SHA-256  & 16         & P-256      & ES256     & AES-CCM-16-64-128 \\[5pt] 
        4                  & ChaCha20/Poly1305  & SHA-256  & 16         & X25519     & EdDSA     & ChaCha20/Poly1305 \\[5pt] 
        5                  & ChaCha20/Poly1305  & SHA-256  & 16         & P-256      & ES256     & ChaCha20/Poly1305 \\[5pt]
        6                  & A128GCM            & SHA-256  & 16         & X25519     & ES256     & A128GCM           \\[5pt] 
        24                 & A256GCM            & SHA-384  & 16         & P-384      & ES384     & A256GCM           \\[5pt]
        25                 & ChaCha20/Poly1305  & SHAKE256 & 16         & X448       & EdDSA     & ChaCha20/Poly1305 \\[5pt]
        -24, -23, -22, -21 & \multicolumn{6}{c}{Private use}                                                         \\[5pt] 
    \end{tabular}
    \caption{Current cipher suites and their identification integer as registered by IANA.}
    \label{tab:ciphersuites}
\end{table*}

To enable crypto agility, flexibility and modularity in the design, EDHOC includes a list of ordered algorithms in its first message.
A single integer encodes the cipher suite: a set of algorithms such as the authenticated encryption algorithm, a hash function, an elliptic curve (see Table~\ref{tab:ciphersuites}).

\paragraph{EDHOC Protocol Outline}

\begin{figure*}
    \centering
     \includegraphics[width=1.00\textwidth]{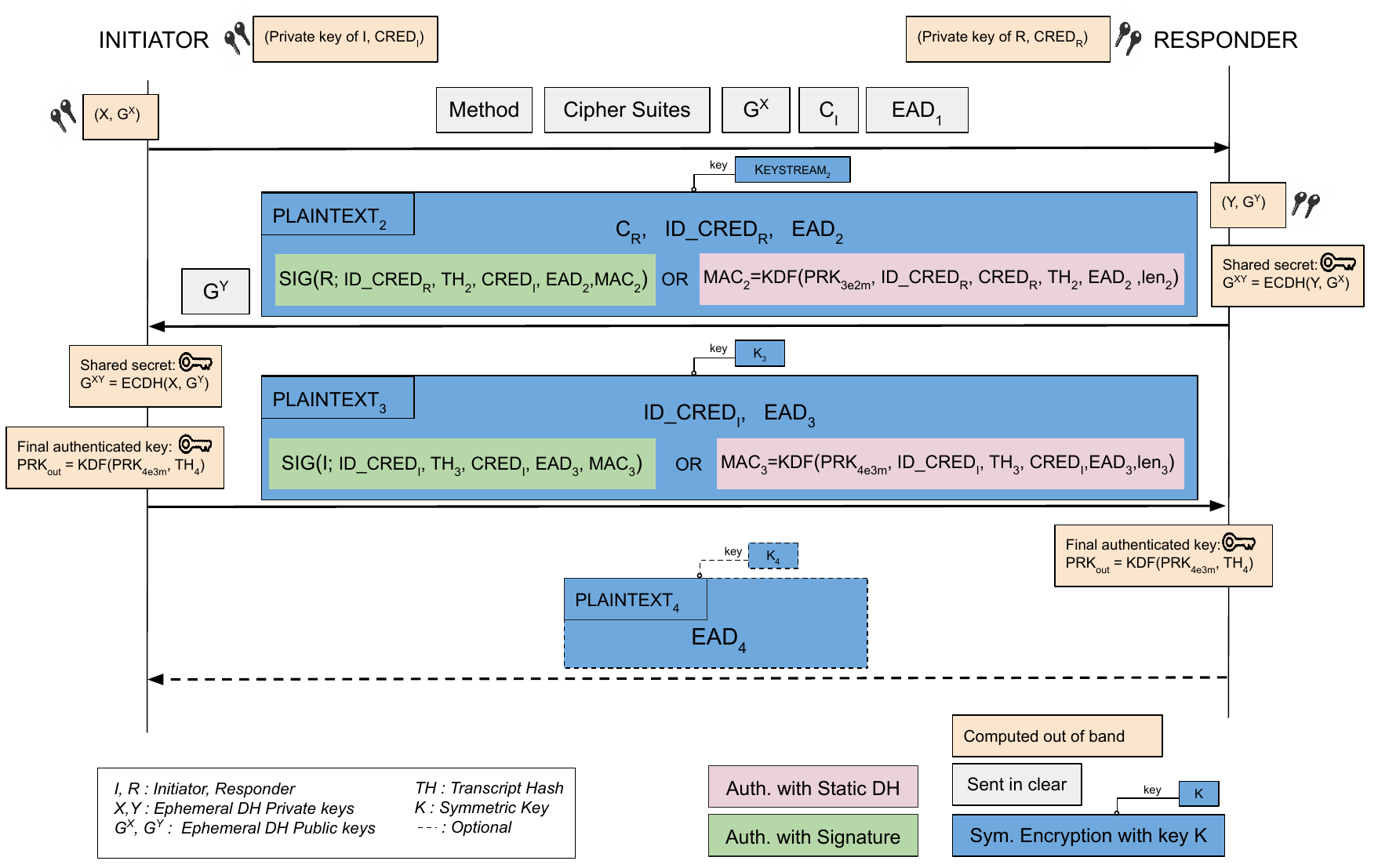}
    \caption{%
        The EDHOC message flow.
        The fields (X, $G^X$) (resp. (Y, $G^Y$)) represent ephemeral private and public key of the Initiator (resp.~Responder).
        Field $CRED_I$ (resp. $CRED_R$) denotes the authentication credentials containing the public authentication keys of I (resp.~R).
        Method is an integer (0-1-2-3) denoting the authentication method (see Table~\ref{tab:method_EDHOC}).
        Cipher Suites is an ordered set of preferred algorithms (see Table~\ref{tab:ciphersuites}).
        If method is either 0 or 1 for the Initiator (resp. 0 or 2 for the Responder), then Sig or MAC equals Sig.
        The fourth message is optional (represented with a dashed line).} 
    \label{fig:EDHOC_message_flow}
\end{figure*}

\begin{figure*}
    \centering
    \includegraphics[width=\textwidth]{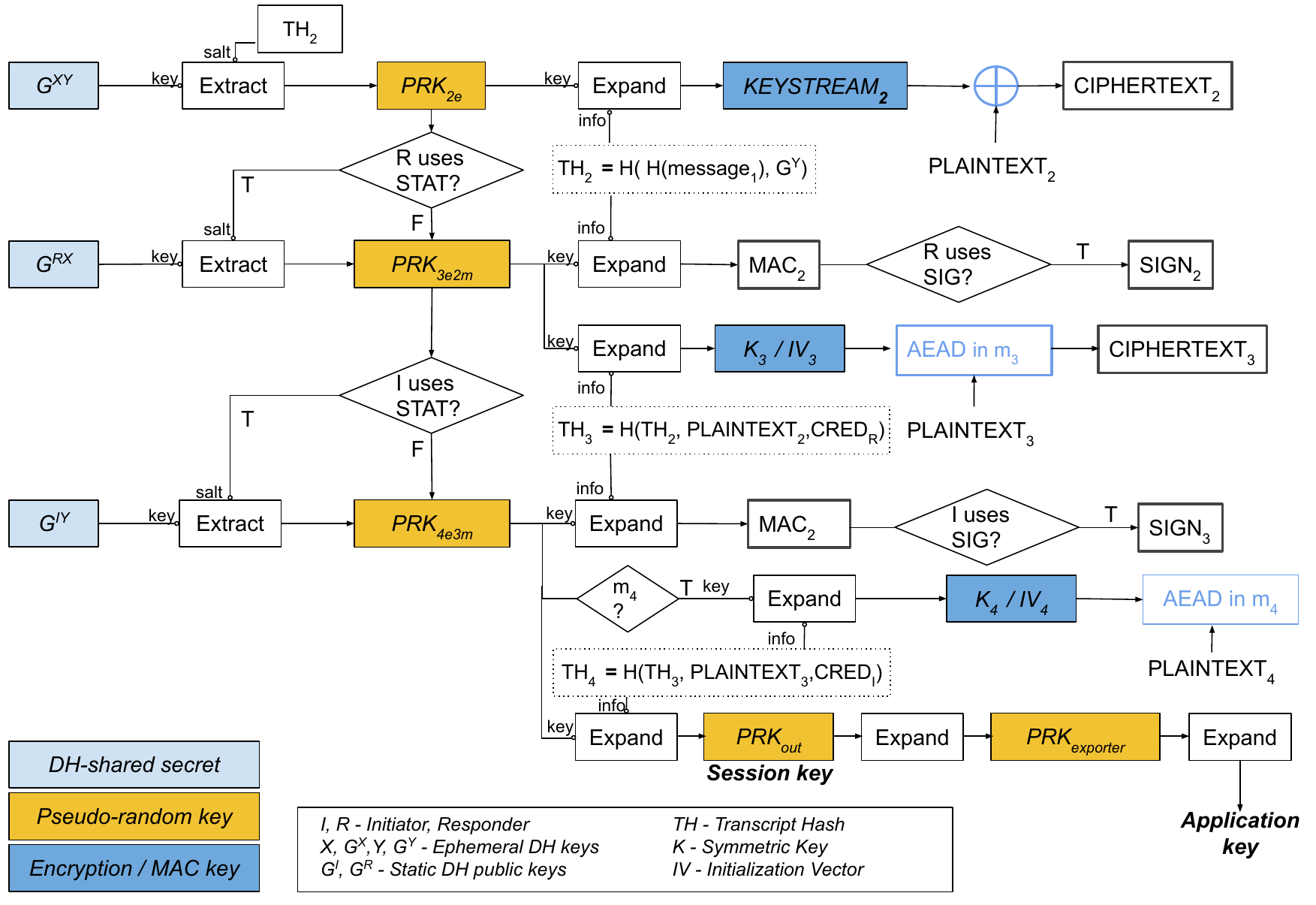}
    \caption{The EDHOC key schedule as standardized in RFC 9528, adapted from Vučinić~\etal~\cite{vucinic22lightweight}.}
    \label{fig:key_shcedule_EDHOC}
\end{figure*}

The protocol consists of three mandatory messages, an optional fourth message, and an error message.
Figure \ref{fig:EDHOC_message_flow} shows the message flow for the EDHOC protocol outlined below.
\begin{itemize}
    \item \textbf{Message~1.}
        The Initiator sends the first message containing the setup information, which includes the authentication method to be used, the cipher suites, the ephemeral Diffie-Hellman public key $G^X$, a connection identifier $C_{I}$ (with no cryptographic purpose) and the additional information called external authorization data $EAD_{1}$. 
    \item \textbf{Message~2.}
        If the Responder agrees on the method and cipher suites, it generates an ephemeral key pair ($Y$, $G^Y$) and computes the Diffie-Hellman shared secret $G^{XY}$.
        If the Responder uses static DH for authentication with key pair ($R$, $G^R$) then it also computes the ephemeral-static Diffie-Hellman shared secret $G^{RX}$.
        The Responder derives two intermediate pseudo-random keys using the Key Derivation Function \texttt{EDHOC\_KDF}: $PRK_{2e}$ and $PRK_{3e2m}$, see Fig.~\ref{fig:key_shcedule_EDHOC}.
        The index in the key name indicates in what message and operation the key is involved. 
        Thus, $PRK_{2e}$ is used for the encryption of the message~2 whereas $PRK_{3e2m}$ is used for encryption of message~3 and MAC of message~2. 
        The Responder also computes a transcript hash $TH_2$ that includes the first message and the ephemeral Diffie-Hellman public key $G^Y$.
        When the Responder uses a Static Diffie-Hellman key, authentication is ensured by a $MAC_2$ (methods 1 and 3).
        This MAC is derived using \texttt{EDHOC\_KDF} with $PRK_{3e2m}$ and some additional information ($context_2$) that includes the previously calculated transcript hash $TH_2$ and its own credentials $CRED_R$.
        Otherwise (methods 0 and 2), the Responder authenticates via a signature.
        The parameters used in the signature algorithm are:
            \textit{(1)} $ID\_CRED_R$, an identifier to facilitate retrieval of $CRED_R$.
            \textit{(2)} external authorization data, including $TH_2$, $CRED_R$ and $EAD_2$, and
            \textit{(3)} the previously calculated $MAC_2$.
        Message~2 is finally encrypted using keystream $K_2$, derived from $PRK_{2e}$.
    \item \textbf{Message~3.}
        After receiving the second message, the Initiator computes the shared secret $G^{XY}$, the decryption key $K_2$ derived from $PRK_{2e}$, and the pseudo-random keys $PRK_{3e2m}$, all used for verifying message 2.
        If the Initiator uses static DH for authentication with key pair ($I$, $G^I$) then it also computes the ephemeral-static Diffie-Hellman shared secret $G^{IY}$ used in the derivation of $PRK_{4e3m}$, see Fig.~\ref{fig:key_shcedule_EDHOC}.
        
        I verifies R's credentials by means of either the signature or MAC. 
        I then computes a new transcript hash $TH_3$ including R's credentials, and a MAC, $MAC_{3}$, using \texttt{EDHOC\_KDF} on $PRK_{4e3m}$ and some additional data ($context_3$) that includes the previous hash $TH_3$ and its own credentials $CRED_I$.
        Depending on the method used, the MAC is either sent raw or protected by a signature. 
        It is further encrypted using keying material $K_3$ / $IV_3$ derived from $PRK_{3e2m}$.
    \item \textbf{Message~4.}
        The Responder receives the third message and decrypts it to authenticate the Initiator. 
        R can then send an optional fourth message containing external authorized data $EAD_4$ as acknowledgment, encrypted using the keying material $K_4$ / $IV_4$ derived from $PRK_{4e3m}$.
\end{itemize}

\paragraph{Key Schedule}

EDHOC uses a similar key schedule to the Noise XX pattern~\cite{perrin18noise},
    a handshake pattern within the Noise protocol framework designed for establishing secure communication channels between two parties with static long-term keys. 

Two functions are involved in the key schedule:
    \textit{(1)} \texttt{EDHOC\_Extract} and
    \textit{(2)} \texttt{EDHOC\_Expand}.
In a first stage, \texttt{EDHOC\_Extract} receives as input a salt and an Input Keying Material (IKM), and generates a fixed-length pseudorandom key (PRK) that is uniformly random-distributed.
In a second stage, \texttt{EDHOC\_Expand} takes the previously generated PRK, a sequence named info, and the key length, and outputs several additional pseudorandom keys that will be used both in the \texttt{EDHOC\_KDF} and as output keying material.
The definition of both functions depends on the hash algorithm of the cipher suites.
Details on the different keys used in EDHOC are illustrated in Fig.~\ref{fig:key_shcedule_EDHOC}.

\section{Security Analysis of EDHOC}

Before TLS~1.3, the process of standardization and formalization often occurred sequentially.
Formalization, which involves the rigorous mathematical analysis of the protocol's security properties, typically occurred independently or as part of academic research efforts. 
Meanwhile, standardization efforts, led by organizations such as the IETF, focused on developing and documenting the protocol specification, considering practical deployment considerations, interoperability requirements, and feedback from implementers.
However, with the increasing recognition of the importance of formal analysis in protocol design, especially in the context of Internet security and privacy, there has been a growing emphasis on integrating formal methods into the standardization process from an early stage. 
In this regard, the standardization of EDHOC followed this parallel approach, with formal analysis informing the design and development of the protocol specification from the outset.

In order to obtain proofs that security protocols are correct, one first needs to model them mathematically. 
Two models of protocols have been considered to study the security requirements of EDHOC:
    \textit{(1)} a symbolic model, and
    \textit{(2)} a computational model.
The symbolic model, often called Dolev-Yao model, is an adversary model that employs idealized cryptographic primitives and that allows the adversary to control the communication channel and interact with protocol sessions by dropping, injecting or modifying messages.
In this model, the cryptographic primitives are represented by function symbols considered as black-boxes, the messages are terms on these primitives, and the adversary is restricted to compute only using these primitives.
As for the computational model, the messages are bit strings, the cryptographic primitives are functions from bit strings to bit strings, and the adversary is any probabilistic Turing machine.
In this model, a security property is considered to hold when the probability that it does not hold is negligible in the security parameter.
In the computational model it is common to use a code-based game framework where initially the problem is represented as a code-based game, named game 0, and it is sequentially modified creating a chain of games in which the last game has the property that it cannot be won by the adversary, i.e.~the advantage is zero.
By using the fundamental lemma of game playing (see Bellare~\etal~\cite{bellare06security} for details), the probability of the adversary winning the original game can be bounded.
Even though the computational model is much more realistic, the symbolic model is suitable for automation, essentially by computing the set of all messages the adversary can know.

\begin{table*}[htbp]
    \centering
    \renewcommand{\arraystretch}{1.5}
    \begin{tabular}{|B{1.7cm}|M{3.5cm}|M{4.5cm}|M{0.8cm}|M{1.2cm}|M{1cm}|M{0.6cm}|M{0.6cm}|}
     \rowcolor{IEEEblue!40}
        \hline
        \textbf{Security goal} & \textbf{Vulnerability} & \textbf{Mitigation} & \textbf{Initial draft} & \textbf{Improved draft} & \textbf{Method} & \textbf{Proof type} & \textbf{Ref.} \\
        \hline
        \hline
        
        & Weak final key.  Reuse of the last key-exchange internal key & Final key depending on $PRK_{4x3m}$ and $TH_4$ ($PRK_{out}$) & 12 & 14 & 0-1-2-3 & S & \cite{jacomme23comprehensive} \\ 
        &\cellcolor{IEEEblue!10} Transcript collision attack & \cellcolor{IEEEblue!10}Reorder arguments in the hash function &\cellcolor{IEEEblue!10} 12 & \cellcolor{IEEEblue!10}14 &\cellcolor{IEEEblue!10} 0-1-2-3 &\cellcolor{IEEEblue!10} S &\cellcolor{IEEEblue!10} \cite{jacomme23comprehensive} \\
        & Duplicate Signature Key Selection (identity misbinding attack) & Include full/unique authentication credentials in the hash function. Build transcript hashes based on plaintext
            & 14 & 17 & 0 & C & \cite{guenther23careful} \\ 
        &\cellcolor{IEEEblue!10}Key reuse & \cellcolor{IEEEblue!10}Not to reuse keys across calls of $EDHOC\_Extract$ and $EDCHO\_Expand$ & \cellcolor{IEEEblue!10}14 &\cellcolor{IEEEblue!10}17  & \cellcolor{IEEEblue!10}0 & \cellcolor{IEEEblue!10}C &\cellcolor{IEEEblue!10} \cite{guenther23careful}\\
        \multirow{-10}{*}{Confidentiality} & Salt Collision Attack & Use $TH_2$ as salt in the HKDF Extract function to derive $PRK_{2e}$  & 15 & 16 & 3 & C & \cite{cottier22security} \\ 
             
        \hline
        & \cellcolor{IEEEblue!10}KCI &\cellcolor{IEEEblue!10} Modify the construction of message 3 &\cellcolor{IEEEblue!10} 15 &\cellcolor{IEEEblue!10} - & \cellcolor{IEEEblue!10}3 & \cellcolor{IEEEblue!10}C& \cellcolor{IEEEblue!10}\cite{cottier22security} \\
        & Leaking ephemeral secrets breaks authentication & Entity authentication should only rely on long-term authentication secrets & 12 & 14 & 0-1-2-3 & S & \cite{jacomme23comprehensive} \\ 
        \multirow{-5}{*}{\begin{tabular}[c]{@{}c@{}}Mutual\\authentication\end{tabular}}  & \cellcolor{IEEEblue!10}Injective agreement &\cellcolor{IEEEblue!10} Add a fourth message as an option &\cellcolor{IEEEblue!10} 00 &\cellcolor{IEEEblue!10} Op. & \cellcolor{IEEEblue!10}0-1-2-3 &\cellcolor{IEEEblue!10} S &\cellcolor{IEEEblue!10} \cite{norrman20formal} \\
             
        \hline
        & Initiator impersonation & Include the Initiator identity in the list of trusted identities for the Initiator & 12 & 14 & 0-1-2-3 & S & \cite{jacomme23comprehensive} \\ 
        \multirow{-3}{*}{\begin{tabular}[c]{@{}c@{}}Identity\\protection\end{tabular}}&\cellcolor{IEEEblue!10} Partial privacy disclosure of the Responder's identity & \cellcolor{IEEEblue!10}Authenticate the first message and provide a way to validate the second message &  \cellcolor{IEEEblue!10}07 &\cellcolor{IEEEblue!10} - & \cellcolor{IEEEblue!10}0-1-2-3 &\cellcolor{IEEEblue!10} S &\cellcolor{IEEEblue!10} \cite{kim21scrutinizing} \\
             
        \hline
        & Attacks in $2^{64}$ operations for the Responder & Add a fourth message  & 15 & Op. & 3 & C & \cite{cottier22security} \\
        \multirow{-3}{*}{\begin{tabular}[c]{@{}c@{}}Cryptographic\\strength\end{tabular}}& & & & & & &\\
            
        \hline
        &\cellcolor{IEEEblue!10} AEAD Key/IV reuse & \cellcolor{IEEEblue!10}Do not allow message recomputation from stored data &\cellcolor{IEEEblue!10} 12 &\cellcolor{IEEEblue!10} 14 &\cellcolor{IEEEblue!10} 0-1-2-3 &\cellcolor{IEEEblue!10} S &\cellcolor{IEEEblue!10} \cite{jacomme23comprehensive} \\
        &\cellcolor{IEEEblue!10} &\cellcolor{IEEEblue!10} &\cellcolor{IEEEblue!10} & \cellcolor{IEEEblue!10}&\cellcolor{IEEEblue!10} &\cellcolor{IEEEblue!10} &\cellcolor{IEEEblue!10}\\
        \multirow{-4}{*}{\begin{tabular}[c]{@{}c@{}}Protection\\of external\\data\end{tabular}}   &\cellcolor{IEEEblue!10}&\cellcolor{IEEEblue!10} & \cellcolor{IEEEblue!10}&\cellcolor{IEEEblue!10} &\cellcolor{IEEEblue!10} &\cellcolor{IEEEblue!10} &\cellcolor{IEEEblue!10}\\
            
        \hline
         & Unclear intended use & The Initiator should verify whether the identity of the Responder matches the intended one. & 00 & 05 & 0-1-2-3 & S & \cite{norrman20formal} \\ 
        & \cellcolor{IEEEblue!10}Malleable signatures & \cellcolor{IEEEblue!10}Do not accept low-order points or the identity group element & \cellcolor{IEEEblue!10}12 &\cellcolor{IEEEblue!10} - & \cellcolor{IEEEblue!10}0-1-2-3 & \cellcolor{IEEEblue!10}S & \cellcolor{IEEEblue!10}\cite{jacomme23comprehensive}\\ 
        \multirow{-5}{*}{\begin{tabular}[c]{@{}c@{}}Non-\\repudiation\end{tabular}}  & Sessions sharing the same $PRK\_4e3m$ & Do not accept low-order points or the identity group element & 12 & - & 0-1-2-3 & S & \cite{jacomme23comprehensive}\\
            
        \hline
    \end{tabular}
    \caption{Different vulnerabilities found during the security analysis of EDHOC and their corresponding mitigations. Op. denotes optional, and - denotes not included. S denotes symbolic analysis and C denotes computational analysis.}
    \label{tab:comparison}
\end{table*}

In this section, we discuss the main vulnerabilities that were found by the community in the security analyses of EDHOC and how they were resolved in order to meet the listed security goals.
A summary of the vulnerabilities and their proposed mitigations can be found in Table \ref{tab:comparison}.

\begin{itemize}
    \item \textbf{Confidentiality.}
        Jacomme~\etal~\cite{jacomme23comprehensive} performed a symbolic analysis of draft~12 on all authentication methods and found a flaw in the key derivation that lead  to a weak session key, due to a reuse of the last key exchange internal key ($PRK_{4e3m}$).
        Their proposed fix was to add a final key derivation that would depend on both $PRK_{4e3m}$ and $TH_4$.
        The advantages of this improvement are:
            \textit{(1)} the session key is different from the MAC key,
            \textit{(2)} a key confirmation implies authentication of all the data,
            \textit{(3)} a dishonest party cannot control the final value of the session key and
            \textit{(4)} the final state of the protocol is simplified since only the session key needs to be stored, rather than storing a key and a transcript. 
        The improvement was added in draft 14 and formally verified (in a computational model) by Günther~\etal~\cite{guenther23careful}.

    
        Later on, G\"unther~\etal~\cite{guenther23careful} performed a computational proof of draft 14 using authentication method 0 and suggested including full/unique credentials in the transcript hashes ($TH_3$ and $TH_4$).
        The aim was to strengthen EDHOC against attacks leveraging non unique credentials, such as identity misbinding attacks, where a legitimate but compromised peer manipulates the honest peer so that it becomes unknowingly associated with a third party.
    
        Lastly, Cottier and Pointcheval~\cite{cottier22security} performed a computational proof of draft 15 using method 3 and detected a salt collision attack (an attacker manages to find two different inputs that, when combined with their respective salts, produce the same hash value).
        Indeed, previous to their suggestion, an empty string was used in the derivation of $PRK_{2e}$.
        However, this means that no additional randomness is added to the hashing process. 
        Thus, they suggested replacing the empty string used as salt with $TH_2$ in the derivation of $PRK_{2e}$.
        This proposal was included in draft 16.
    
    \item \textbf{Mutual authentication.}
        Norrman~\etal~\cite{norrman20formal} performed a formal analysis of all four authentication methods in draft~00 in a symbolic model. 
        In their analysis, they proved two flavors of mutual authentication.
        The first one is injective agreement, which guarantees to an initiator I that whenever I completes a run with a Responder R, then R has been engaged in the protocol as a Responder.
        It also guarantees that I and R agree on a set of parameters, including the session key material.
        The corresponding property for R is analogous.
        The second variant of authentication is implicit agreement, which establishes that if the Initiator I (resp. Responder R) assumes that the Responder R (resp. Initiator T) knows the session key material, then it must be the intended party.
        In their analysis they show that implicit agreement on the session key material and the initiator's identity hold for all four authentication methods.
        Contrary, given that EDHOC aims to protect the Initiator's identity, injective agreement on the initiator's identity does not hold for any of the methods.
        As for the injective agreement on the session key material, it is not supported whenever the Initiator uses methods 1 or 3, i.e., static Diffie-Hellman key.
        To solve it, they propose to add a fourth message from R to I carrying a MAC based on a key derived from the session key material.
        The optional message four was added by the LAKE Working Group as an optional feature of the protocol.

        Cottier and Pointcheval~\cite{cottier22security} completed a computational proof of draft 15 using authentication method 3, i.e., static Diffie-Hellman keys.
        They discovered a Key Compromise Impersonation (KCI) attack on $K_3$, where an attacker gains access to a peer's private key and compromises and impersonates this peer.
        It turns out that $K_3$, used by the Initiator to encrypt message~3, is computed by calling \texttt{EDHOC$\_$Expand} on $PRK_{3e2m}$, and can be computed by an adversary that knows the Initiator ephemeral key $G^X$.
        In order to break the Initiator's authentication, an adversary must know as well $MAC_{3}$.
        However, in the most constrained cipher suites (identifiers 0 and 2 in Table~\ref{tab:ciphersuites}), $MAC_{3}$ is 64~bits long.
        They suggest modifying the construction of message~3 by splitting it in two parts:
            \textit{(1)} the first one contains the identity credentials of the initiator and is encrypted using the corresponding encryption key $K_3$ derived from $PRK_{3e2m}$,
            \textit{(2)} the other half corresponds to the authentication tag $MAC_{3}$ and the $EAD_{3}$ and is sent without encryption.
        The main reason why this change was not included in the draft was that it did not offer protection against active attackers, since the identity credentials were not be protected with a MAC. 
        The proposed alternative is to use a longer 16 bytes $MAC_{3}$, which was already contemplated by some cipher suites, such as 1 and 3 (see Table~\ref{tab:ciphersuites}).

        Ferreira~\cite{ferreira24computational} performed a computational analysis of the final draft of EDHOC, draft 23, on the four authentication methods. 
        Similarly to Norrman~\etal~\cite{norrman20formal}, they proved that a fourth message is necessary only when the initiator authenticates by means of a static DH key, i.e.~authentication methods~1 and~3.
        For these authentication methods, the fourth message proves to the initiator that the responder shares the same key $PRK_4$.
        Furthermore, it also proves to the initiator that the responder shares the same session identifier.

    \item \textbf{Identity protection.}
        Kim~\etal~\cite{kim21scrutinizing} studied draft 7 leveraging BAN Logic, a set of rules for defining and analyzing information exchange protocols, and AVISPA (Automated Validation of Internet Security Protocols and Applications), a tool designed for the automated analysis and validation of security protocols and applications.
        Their results show vulnerability against privacy attacks affecting the identity credentials ($ID\_CRED_R$ and $ID\_CRED_I$).
        Thus, an attacker can easily break the privacy of $ID\_CRED_R$ by establishing the session key with the Responder, leading to a partial privacy disclosure of the Responder's identity.
        They recommend authenticating message 1 and providing a way to validate message~2.
    
        Jacomme~\etal~\cite{jacomme23comprehensive} detected a privacy leak in draft 12 when the list of trusted identities for the Initiator I is only missing I's identity.
        They proposed to fix it by adding the Initiator identity in its own list of trusted identities.
        
        In their computational analysis of draft~23 of the protocol, Ferreira~\cite{ferreira24computational} suggests that identity privacy can be enhanced by systematically using padding before encryption, particularly with plaintext elements like $EAD_2$, $EAD_3$ and $EAD_4$.
        Padding can obscure the actual size of the data being transmitted, making it harder for an adversary to infer information about the identities involved.
        However, padding should be used with care as it enlarges the messages and increases bandwidth consumption.

    \item \textbf{Cryptographic strength.}
        The cryptographic strength of the protocol was formally studied by Cottier and Pointcheval~\cite{cottier22security}.
        They proved that in the three-flow scenario and using the most constrained cipher suites with 8-byte long MACs, the protocol provides a 64-bit security level for the Responder authentication.
        They further showed that a fourth message using authenticated encryption (AEAD) from the Responder to the Initiator increases this security up to a 128-bit level, being consistent with the security requirements.

        Ferreira~\cite{ferreira24computational} showed that the security of EDHOC depends essentially on that of the authenticated encryption algorithm used.
        They show that the cipher suite based on A256GCM is preferable for high security applications, in contrast with ChaCha20Poly1305, also recommended in the EDHOC draft for the same kind of applications.

    \item \textbf{Protection of external data.}
        Jacomme~\etal~\cite{jacomme23comprehensive} detected an attack in draft 12 due to the possibility of resending the last message, which lead to a nonce reuse of AEAD, breaking confidentiality and integrity.
        They therefore suggested not to allow message resending when using randomized signatures.
        This was specified in draft 14.

    \item \textbf{Non-repudiation.}
        Non-repudiation was highlighted by Norrman~\etal~\cite{norrman20formal} in their analysis of draft~00 and Jacomme~\etal~\cite{jacomme23comprehensive} in their study of draft~12. 
        They signaled cases in which non-repudiation would be unclear.
        These include:
            \textit{(1)} sessions sharing the same $PRK_{4e3m}$ due to acceptance of low-order points or the identity group element and
            \textit{(2)} signatures being malleable.
        These situations were considered side cases and were thus not included by the Working Group in the draft.

    \item \textbf{Downgrade protection.}
        There were no vulnerabilities found regarding downgrade attacks in any of the studies. 
        The defined cipher suite negotiation in the specifications is thus found to be resistant against these attacks.
\end{itemize}

\begin{table}
    \centering
    \begin{tabular}{c|c|c|c|c}
        \rowcolor{IEEEblue!40}
        \textbf{Security goal} & \multicolumn{4}{c}{\textbf{Method}} \\[5pt] \cline{2-5}
         \cellcolor{IEEEblue!40}& \cellcolor{IEEEblue!40} 0 & \cellcolor{IEEEblue!40}1 & \cellcolor{IEEEblue!40}2 & \cellcolor{IEEEblue!40}3 \\
        \hline
        \cellcolor{white}Confidentiality & \cellcolor{white}S,C & \cellcolor{white}S,C & \cellcolor{white}S,C & \cellcolor{white}S,C \\[5pt]
       \cellcolor{IEEEblue!10}Mutual authentication & \cellcolor{IEEEblue!10}S,C &  \cellcolor{IEEEblue!10}S,C & \cellcolor{IEEEblue!10}S,C & \cellcolor{IEEEblue!10}S,C \\[5pt]
       \cellcolor{white}Identity protection & \cellcolor{white}S,C & \cellcolor{white}S,C &\cellcolor{white}S,C &\cellcolor{white}S,C \\[5pt]
        \cellcolor{IEEEblue!10}Cryptographic strength & \cellcolor{IEEEblue!10}C &  \cellcolor{IEEEblue!10}C & \cellcolor{IEEEblue!10}C & \cellcolor{IEEEblue!10}C \\[5pt]
       \cellcolor{white}Protection of external data & \cellcolor{white}S & \cellcolor{white}S &\cellcolor{white}S &\cellcolor{white}S \\[5pt]
        \cellcolor{IEEEblue!10}Downgrade protection & \cellcolor{IEEEblue!10}S & \cellcolor{IEEEblue!10}S &\cellcolor{IEEEblue!10}S &\cellcolor{IEEEblue!10}S \\[5pt]
    \end{tabular}
    \caption{Summary of the EDHOC security analyses. \textit{S}~denotes symbolic proof and \textit{C} computational proof. Method refers to the authentication method (listed in Table \ref{tab:method_EDHOC}).}
    \label{tab:security_goals}
\end{table}

Table~\ref{tab:security_goals} summarizes how EDHOC meets the discussed security goals, including the corresponding authentication method for which the proof holds and the type of proof.

\section{Discussion}


The security analyses of EDHOC are restricted to the cryptographic core of the protocol, and thus, they do not capture all aspects of the protocol like negotiation of authentication methods and cipher suites.

Moreover, they do not consider other attack surfaces in low-power devices that may undermine the security guarantee of EDHOC.
Some examples include side-channel attacks (an adversary exploits information leak from the physical implementation of cryptographic operations, such as power consumption or electromagnetic emissions) or fault injection attacks (an adversary introduces faults or errors into a system's operation with the aim of compromising its security or integrity).


EDHOC has included a Post Quantum Cryptography (PQC) variant that employs a PQC Key Encapsulation Method (KEM).
Even though EDHOC is currently only specified for use with key exchange algorithms of type ECDH curves, any KEM, including PQC KEMs, can be used in method 0.
However, current PQC algorithms have limitations compared to Elliptic Curve Cryptography (ECC), and the data sizes would be problematic in many constrained IoT systems.
The use of other key exchange algorithms to replace static Diffie-Hellman authentication (methods 1,2 and 3) will require the definition of new methods in the specification.
Up to date, there are no security analyses of the Post Quantum variant.
Even though Jacomme~\etal~\cite{jacomme23comprehensive} mentioned in their work that they studied it, no results were presented. 
Given the evolution of quantum computers, it would be interesting to perform further analyses in the future.


Lastly, the LAKE Working Group plans to work on a new pre-shared key (PSK)-based method, to be also employed for rekeying, which will require a formal analysis in the future.

\section{Conclusion}


The rapid growth of Internet of Things (IoT) calls out for developing lightweight and efficient security protocols that meet the particular IoT requirements, such as scarce bandwidth, low processing power or limited battery.
In this regard, the Lightweight Authenticated Key Exchange (LAKE) working group developed the Ephemeral Diffie-Hellman Over COSE (EDHOC) protocol, an authenticated key exchange protocol for constrained environments.
Despite the existence of TLS and DTLS, EDHOC proves to be better suited for resource constrained devices due to its reduced message footprint, the limited number of flights (3 and optionally 4), the fact that is transport agnostic and the reduced code size (because of reusing OSCORE elements).


The Working Group has solicited formal analysis from the community to incorporate feedback and improvements.
The formalization and standardization of EDHOC have been done in parallel, following the approach used in TLS 1.3.
This paper summarizes the different studies that have been performed since draft 00 (July 2020) until standard has been published (March 2024), as well as the changes that have been incorporated as a result of them.
The different analyses include both symbolic and computational proofs, and they comprise different authentication methods.
We point out the main vulnerabilities that have been found, what were the proposals to fix them and how they contribute to EDHOC meeting the security requirements established by Vučinić~\etal~\cite{draft-ietf-lake-reqs-04}.

In the paper we also present the current state of the protocol, including security requirements, message flow and key scheduling.

\section{Acknowledgments}
This work has been partially supported by the French National Research Agency under the France 2030 label (NF-HiSec ANR-22-PEFT-0009) and by the Horizon Europe OpenSwarm project under Grant Agreement No. 101093046.
The views reflected herein do not necessarily reflect the opinion of the French government nor of the European Commission.
The authors would like to thank the participants of the IETF LAKE Working Group for their input during the standardization process.

\bibliographystyle{unsrt}
\bibliography{lopezperez24edhoc}

\begin{IEEEbiography}{Elsa López Pérez}{\,}%
is a PhD student at Inria, Paris, France.
She received her double master degree in Data Science and Artificial Intelligence from University of Paris Saclay and University of Trento.
Contact her at elsa.lopez-perez@inria.fr.
\end{IEEEbiography}

\begin{IEEEbiography}{Göran Selander}{\,}%
is a principal security researcher at Ericsson Research, Stockholm, Sweden.
Contact him at goran.selander@ericsson.com.
\end{IEEEbiography}

\begin{IEEEbiography}{John Preuß Mattsson}{\,}%
is a senior security expert in cryptographic algorithms and security protocols at Ericsson Research, Stockholm, Sweden.
Contact him at john.mattsson@ericsson.com.
\end{IEEEbiography}

\begin{IEEEbiography}{Thomas Watteyne}{\,}%
is a research director at Inria, Paris, France.
Contact him at thomas.watteyne@inria.fr.
\end{IEEEbiography}

\begin{IEEEbiography}{Mališa Vučinić}{\,}%
holds a starting faculty position at Inria, Paris, France.
Contact him at malisa.vucinic@inria.fr.
\end{IEEEbiography}

\end{document}